\documentstyle{mn}

\def\nhi{\mbox{$N_{\rm HI}$}}

\def\q0{q$_0$}
\def\lya{Ly$\alpha$\ }

\def\etal{\rm et al.}
\def\eg{\protect\rm e.g.}
\def\ie{\protect\rm i.e.}
\def\cf{\protect\rm cf.}

\def\simlt{$_<\atop{^\sim}$}
\def\simgt{$_>\atop{^\sim}$}

\def\vs{\vspace{1ex}}
\def\notes{\vs\hspace{10mm}}
\def\hi{\mbox{\rm HI}}

\def\cii{\mbox{\rm CII}}
\def\oi{\mbox{\rm OI}}
\def\o1{\mbox{\rm OI}}
\def\si2{\mbox{\rm SiII}}

\def\os2{\rm OI+SiII}
\def\lyb{Ly$\beta$}

\def\ovi{\rm OVI}

\def \m.        {\rlap{$.$}^{\rm m}}
\def \s.        {\rlap{$.$}^{\rm s}}
\def \am.       {\rlap{$.$}'}
\def \as.       {\rlap{$.$}''}

\title[APM z \simgt\ 4 QSO Survey ... High Column Density HI Absorbers]
{ APM z \simgt\ 4 QSO Survey: 
Distribution and Evolution of High Column Density HI Absorbers}
\author[L.J. Storrie-Lombardi, M.J. Irwin, R.G. McMahon ]
{L. J. Storrie-Lombardi$^{1,2}$, M. J. Irwin$^3$, 
\& R. G. McMahon$^1$ \\
$^1$Institute of Astronomy Madingley Road, Cambridge CB3 0HA, UK \\
$^2$current address: Carnegie Observatories, 813 Santa Barbara Street,
Pasadena, CA  91101 USA \\ 
$^3$Royal Greenwich Observatory Madingley Rd, Cambridge CB3 0EZ, UK \\
email:  lisa@ociw.edu, mike@ast.cam.ac.uk, rgm@ast.cam.ac.uk \\ \ \\
in press}

\begin{document}

\maketitle

\begin{abstract}

Eleven candidate damped \lya absorption systems
were identified in twenty-seven spectra of the quasars from the 
APM z \simgt\ 4 survey covering
the redshift range $2.8\le z_{absorption} \le 4.4$ (8 with $z_{absorption}>3.5$).
High resolution echelle spectra (0.8\AA\ FWHM) have been
obtained for three quasars, including two of the highest 
redshift objects in the survey.  
Two damped systems have confirmed HI column densities
of \nhi $\ge$ 10$^{20.3}$ atoms cm$^{-2}$, with a third
falling just below this threshold.
We have discovered the highest redshift damped \lya 
absorber known at z$=$4.383 in QSO BR1202$-$0725. 

The APM QSOs provide a substantial increase in the redshift path
available for damped surveys for z$>$3.
We combine this high redshift sample with other 
quasar samples covering the redshift range 0.008 $<$ z $<$ 4.7 to 
study the redshift evolution and the column density 
distribution function for absorbers with log \nhi$\ge17.2$.
In the \hi\ column density distribution $f(N)=kN^{-\beta}$
we find evidence for breaks in the power law, flattening
for $17.2\le\log$\nhi\simlt21 and steepening for 
$\log$\nhi$>21.2$.  The breaks are more pronounced at 
higher redshift. The column density distribution function 
for the data with log \nhi$\ge$20.3 is
better fit with the form 
$f(N)=(f_*/N_*)(N/N_*)^{-\beta}exp(-N/N_*)$ with 
log $N_*=21.63\pm0.35$,
$\beta=1.48\pm0.30$, and $f_*=1.77\times 10^{-2}$.
We have studied the evolution of the 
number density per unit redshift of the damped systems by
fitting the sample with the customary power law $N(z)=N_0(1+z)^{\gamma}$.
For a population with no intrinsic evolution
in the product of the absorption cross-section and comoving spatial
number density this will give $\gamma=1/2$ ($\Omega = 1$)
or $\gamma=1$ ($\Omega=0$).
The best maximum likelihood fit for a single power law 
is $\gamma=1.3\pm0.5$ and $N_0=.04^{+.03}_{-.02}$,
consistent with no intrinsic evolution
even though the value of $\gamma$ is also consistent 
with that found for the Lyman limit
systems where evolution is detected at a significant level.
However, redshift evolution is evident in the
higher column density systems with an apparent decline in $N(z)$
for z$>$3.5.
\end{abstract}


\section{Introduction}

This paper is the third of a series presenting results from
studies of the QSOs discovered in the APM survey for z \simgt\ 4
quasars.  A study of the evolution of Lyman limit absorption 
systems over the redshift range 0.04 $\le$ z $\le$ 4.7 
was presented in Storrie-Lombardi \etal\ (1994) [Paper I]. 
The intermediate resolution (5\AA) QSO spectra and the survey 
for high redshift damped \lya absorbers are presented in 
Storrie-Lombardi \etal\ (1996) [Paper II].  The evolution of the 
cosmological mass density of neutral gas
at high redshift and the implications for galaxy formation 
theories are discussed in Storrie-Lombardi, McMahon \& Irwin
(1996) [Paper IV].
In separate papers we will describe the intrinsic properties
of the QSOs and  studies of the \lya forest clouds at high redshift.
A high resolution study of the \lya forest region 
in a redshift $z=4.5$ QSO has been completed by Williger \etal\ (1994).

How and when galaxies formed are questions at the forefront
of work in observational cosmology.  
Absorption systems detected in quasar spectra 
provide the means to study galaxy formation and evolution 
up to redshifts of 
approximately five, back to when the Universe was less than 10 percent
of its present age. 
Surveys for absorption features  
have several advantages over trying to directly detect galaxies 
at high redshift.  Much shorter exposure times 
are required because the QSOs are relatively bright 
(R$\approx$ 18-19.5) and the large equivalent width systems are
easily detected in the spectra.  This provides good 
absorption candidates to follow up with higher resolution 
spectra. The redshift and column density can be accurately 
determined from the wavelength of the absorption system
and the line profile.
This is far easier and more reliable than trying to directly get 
a spectrum of a very faint high redshift galaxy.  

While the baryonic content of spiral galaxies 
that are observed in the present epoch is concentrated in stars, 
in the past this must have been in the form of gas. The principal gaseous
component in spirals is neutral hydrogen which has led
to surveys for absorbers detected by the damped \lya
lines they produce (Wolfe \etal\ 1986, hereafter WTSC; 
Lanzetta \etal\ 1991, hereafter LWTLMH; 
Lanzetta, Wolfe \& Turnshek 1995, hereafter LWT; 
Wolfe \etal\ 1995; Paper II). 
Though damped \lya systems are observationally very rare objects 
with $\sim$40 confirmed examples known, the
\hi\ mass per unit comoving volume they
contain is roughly comparable to the
mass density of baryonic matter in present-day spirals, ie. a major
constituent of the Universe (Wolfe 1987, LWT).  
Their metal abundances are much lower than Galactic 
values (Pettini, Boksenberg \& Hunstead 1990; Rauch \etal\ 1990; 
Pettini \etal\ 1994)
and they are characterised by low molecular content
and low, but not negligible, dust content 
(Fall, Pei \& McMahon 1989; Pei, Fall \& Bechtold 1991; 
Pettini \etal\ 1994), 
features consistent with an early phase of galactic evolution.
They may be the progenitors of  
spiral galaxies like our own and are clearly important for the 
study of the formation and evolution of galaxies. 
They have been detected across a very large redshift range
z$\approx$[0.5,4.5] providing the means to pinpoint
the epoch of formation of disk galaxies and study their evolution.

Eleven candidate damped \lya absorption systems
out of 32 measured \lya features were identified in 
27 spectra of the mainly non-BAL quasars from the
APM z \simgt\ 4 survey (Paper II). The eleven candidates
cover the redshift range $2.8\le z_{absorption} \le 4.4$ 
(8 with $z_{absorption}>3.5$)
and have estimated column densities
\nhi $\ge$ 10$^{20.3}$ atoms cm$^{-2}$. 
In this paper the QSO BR1144$-$0723 with a candidate 
absorber at z$=$3.26 is removed from further consideration 
in the sample. 
It has been observed with the Anglo-Australian Telescope
at high resolution and the damped candidate has been found to
be all \ovi\ absorption at z$=$4.0 (R. Hunstead, private communication).
High resolution echelle spectra (0.8\AA\ FWHM) were obtained 
by S. D'Odorico as part of the ESO key programme 
studying high redshift quasars 
for four of the QSOs in the APM sample
(BRI0952$-$0115, BR1033$-$0327, BRI1108$-$0747, BR1202$-$0725).
The signal-to-noise ratio in BRI0952$-$0115 was very poor
but the other spectra have been used to confirm two \lya features 
as damped with another falling just below the log \nhi $\ge$ 20.3
threshold.
We have discovered the highest redshift damped \lya 
absorber known at z$=$4.383 in QSO  
BR1202$-$0725. 
The confirmation of the absorption systems is discussed in section 2.
These data have been combined with data from previous surveys
(WTSC, LWTLMH, and LWT) and the results for the Lyman limit systems
obtained in Paper I to study the \hi\ column density distribution 
for $\log$\nhi $>$ 17.2 and redshift evolution of these systems  
for 0.008 $<$ z $<$ 4.7.

Numerous authors have studied the distribution of column densities, 
$f({\nhi})$,
for \lya absorption lines.  The first determination was by 
Carswell \etal\ (1984) for lines with $10^{13} <$ \nhi $< 10^{16}$ 
atoms cm$^{-2}$.
They found $f(N) \propto N^{-\beta}$ ($\beta=1.7\pm0.1$). 
Damped \lya absorption (DLA) systems comprise the 
high column density tail of neutral hydrogen absorbers  
with column densities of  
\nhi $\ge 2 \times 10^{20}$ atoms cm$^{-2}$. 
They dominate the baryonic mass contributed by \hi.  
When damped systems are included in the column density distribution
function for a single power law fit the exponent is  
$\beta$ = 1.4--1.7 (\cf\ Tytler 1987; 
Petitjean \etal\ 1993 and references therein).
Assuming the baryonic mass is proportional
to the \hi\ column density and takes the form $f({\nhi})\propto N^{-\beta}$
for the \hi\ column density distribution function, the  
mass contribution from the damped systems can be estimated as 
\begin{eqnarray*}
M_{total} &\propto& \int_{N_1}^{N_2} \nhi f({\nhi}) d\nhi \\
   &\propto& \int_{N_1}^{N_2} N^{-\beta} N dN { \ \ \ \ \ (\rm assume\ } \beta \ne 2) \\
   &\propto& {1 \over 2 - \beta} \Bigl[ N_2^{2-\beta} - N_1^{2-\beta} \Bigr].
\end{eqnarray*}
\begin{equation}
\label{himass1}
\end{equation}
One problem with the power law representation is that if $\beta < 2$, 
as all current estimates indicate, then the total mass in damped systems 
diverges unless an upper bound to the HI column density is assumed.
For example, if we take 20.3 \simlt\ log $N_{\rm HI}$ \simlt\ 22, the
fractional contribution to the total HI mass for damped systems, $M_f$,
is then $M_f = 0.86$ for $\beta = 1.5$ and $M_f = 0.69$ for $\beta = 1.7$.
However, there is no a priori reason for assuming this upper limit and 
hence there is no strict upper bound to any estimate of the total HI
mass in damped systems.  An alternative parameterisation using a gamma 
function to describe the HI column density distribution was adopted by
Pei \& Fall \shortcite{PF95} and provides an elegant solution 
to the diverging mass 
problem.  
We discuss these points in more detail in section 3
and the redshift evolution of the absorbers in section 4.

\section{Confirmation of Damped \lya Systems}

\subsection{Echelle Observations}

Echelle spectra of four QSOs were obtained in March,
1993 by S. D'Odorico as part of an ESO 
key programme studying high redshift quasars. They were taken
at La Silla with the 3.5m NTT telescope using the EMMI instrument
in echelle mode using a 2048$\times$2048 pixel 
LORAL CCD as the detector. A slit of 15" in length was used 
and generally the slit width 
was 1.2".  Two grating setups were used, one covering
4700-8300\AA\ and the other covering 5800-9500\AA\ with a 
resolution of $\sim$40 km s$^{-1}$ (1\AA).
[See Giallongo \etal\ 1994 for more details.]  
The observations are summarised in table~\ref{t_obs_eso}.

\begin{table}
\caption{ESO Observations March, 1993}
\label{t_obs_eso}
\begin{tabular}{llrll}
\hline
QSO & Date (UT) & Exp & Grating & Slit \\
    &           & (secs) & & (")\\
\hline
BRI$0952-0115$ & 93 Mar 14 &  5400 &  GR9CD3 &  1.2 \\
              &  93 Mar 14 &  7200 &  GR9CD3 &  2.0 \\
BR $1033-0327$ & 93 Mar 15 &  7200, 8000 &  GR9CD3 & 1.2 \\
BRI$1108-0747$ & 93 Mar 15 &  6000 &  GR9CD3 &  1.2  \\
               & 93 Mar 15 &  6000 &  GR9CD4 &  1.2  \\
BR $1202-0725$ & 93 Mar 14 &  8000 &  GR9CD3 &  1.2  \\
               & 93 Mar 15 &  6797 &  GR9CD3 &  1.2  \\
               & 93 Mar 15 &  7200, 8000 &  GR9CD4 &  1.2 \\
\hline
\end{tabular}
\end{table}

The spectra were extracted 
and calibrated using the OPTEXT routines\footnote{
These were written by Bob Carswell, Jack Baldwin,
and Gerry Williger for the reduction of CTIO echelle data.}
in conjunction with IRAF.
The final flux calibrated spectra agreed well with the 
existing 5\AA\ spectra (Paper II) except in the case of 
BRI0952$-$0115 where the signal-to-noise ratio was very poor
and the relative flux different by a factor of two. This
spectrum has been excluded from further analysis.
Though not of exceedingly high quality
the spectra were suitable for fitting the damped \lya candidates.

\subsection{Fitting the Damped Candidates}

The damped candidate system profiles were fit using the 
{\bf vpgti}\footnote{
Written by R.F. Carswell, J.K. Webb, A.J. Cooke, and M.J. Irwin, see 
also Cooke (1994).}
Voigt profile fitting software. 
The programme requires three input spectra:
the object, the errors, and the continuum. The OPTEXT routines
used in the reduction created the first two and the continuum 
spectra were created by using continuum fits 
for the WHT spectra (Paper II) and extrapolating them 
to the echelle spectra as was done in Williger \etal\ 1994.  
The centroids of the \lya features were determined 
from narrow metal lines (\eg\ OI, CII) and then the
\lya lines were fit in a region 50-100\AA\ around the 
candidate feature.  
Generally, candidate damped systems
seen in the low resolution spectra appeared much stronger than they
actually are, due to blending of the dense Lya forest lines at low
resolution.  In all three QSOs the candidate damped systems are seen
at higher resolution to break up into multiple overlapping components.
The complexity of the blend coupled with residual uncertainty in the
exact placing of the continuum level causes some of the error estimates
from VPFIT to be somewhat optimistic.  In particular for BRI1108$-$0747 and
BR1202$-$0725 numerical simulations suggest an error in log 
\nhi\ of $\sigma=0.15$ should be adopted.
The fitted lines are summarised in table~\ref{t_dla_stats}.

\notes a) BR1033$-$0327, ($z_{em} = 4.509$, $z_{absorption} = 4.15$)
The absorption candidate at z$=$4.15 had been previously studied 
\cite{Williger94} with a 12 km s$^{-1}$ echelle spectrum
taken at CTIO.  Their spectrum covered only the blue wing
of the system and from this the column density was estimated to 
be no greater than $3\times10^{20}$ atoms cm$^{-1}$.
With spectral coverage of the entire absorber it is seen 
to be a complex system of at least 5 absorbers with a total
column density of log \nhi$=20.15\pm0.11$ atoms cm$^{-2}$
($1.4 \times 10^{20}$).
In Paper II the column density was estimated to be log \nhi$=20.2$. 
The redshifts of the \lya absorbers were determined from 
5 tentatively identified \oi 1302 lines in the CTIO spectrum.
The signal-to-noise ratio in the
order where the \lyb\ lines lie was too low to use in the fit.  
The spectrum with the best fit line profiles and 
$\pm 1 \sigma$ fits are shown as solid lines in figure~\ref{f_dla}(a). 
The error array and normalised continuum are shown as dotted and 
dashed lines, respectively. 
The small spike at $\approx$6270\AA\ is real.  It also appears
in the 5\AA\ resolution spectrum (Paper II).

\notes b) BRI1108$-$0747, ($z_{em} = 3.922$, $z_{dla} = 3.607$) 
The absorber at z$=$3.61 is barely damped
with a column density of $\log$ \nhi$=20.33\pm0.15$
atoms cm$^{-2}$. In Paper II we estimated the column density to
be $\log$ \nhi$=20.2$, so this system was not originally in 
the statistical sample. 
Several of the absorbers in the survey in Paper II 
have estimated column densities near the statistical sample threshold
of $\log$ \nhi$=20.3$. We expect some to be confirmed above this
value and some below.   
The profile fit and $\pm 1 \sigma$ fits 
are shown in figure~\ref{f_dla}(b).
The redshift centroid was determined from a single strong
\cii 1334 line.  

\notes c) BR1202$-$0725, ($z_{em} = 4.694$, $z_{dla} = 4.383$)
The damped system in this QSO has a column density of 
$\log$ \nhi$=20.49\pm0.15$.  
It is the highest redshift damped \lya system
known. The ESO spectrum is shown with the profile fit
and $\pm 1 \sigma$ fits in figure~\ref{f_dla}(c).
It was also measured in a higher
resolution spectrum taken at CTIO \cite{Wampler96}, and includes 
over 10 components. Lu {\etal} (1996) have observed it at Keck
with HIRES and measure a column density of $\log$ \nhi$=20.6\pm0.1$. 
It was estimated in Paper II to have a column 
density of $\log$ \nhi$=20.5$.
\begin{table}
\caption{Lyman-$\alpha$ Absorption Systems}
\label{t_dla_stats}
\begin{tabular}{lll}
\hline
QSO & \multicolumn{1}{c}{Absorption}& \multicolumn{1}{c}{log \nhi}\\
    & \multicolumn{1}{c}{Redshift}  &         \\
\hline
BR$1033-0327$ & 4.14945 & 19.80 $\pm$ 0.10  \\
              & 4.15314 & 18.69 $\pm$ 0.30  \\
              & 4.16647 & 19.70 $\pm$ 0.15  \\
              & 4.16726 & 19.37 $\pm$ 0.42  \\
              & 4.17481 & 19.60 $\pm$ 0.24  \\
 & &  \\
BRI$1108-0747$ & 3.60673  & 20.33 $\pm$ 0.15     \\
 & &  \\
BR$1202-0725$  & 4.38290  & 20.49  $\pm$ 0.15  \\
\hline
\end{tabular}
\end{table}
The status of the systems 
detected in the APM Damped \lya absorption survey 
is summarised in table~\ref{t_dla}.
(It is an updated version of table 7 from paper II.)
Those marked with an asterisk have column densities 
log \nhi $\ge$ 20.3 atoms cm$^{-2}$ and make up the statistical 
sample of high redshift absorbers used in the analysis. 
In Paper IV this data set is combined with previous surveys to study the 
evolution of the cosmological mass density of neutral gas
at high redshift and the implications for galaxy formation 
theories are discussed. 
\onecolumn
\begin{table}
\caption{Status of Absorbers in APM Damped Lyman-$\alpha$ Absorption Survey}
\label{t_dla}
\begin{tabular}{lllllrrl}
\hline
QSO & z$_{min}$ & z$_{max}$ & z$_{em}$ & z$_{dla}$ & \multicolumn{1}{c}{W$_{\rm rest}$} &
 \multicolumn{2}{c}{log \nhi}\\
 & &  &  &  & \multicolumn{1}{c}{\AA} & \multicolumn{2}{l}{estimate \ \ \ actual} \\
\hline
BR $0019-1522$ & 2.97 & 4.473 & 4.528 & 3.42  & 7.6 & 20.0 & \\
& &  &   & 3.98 & 12.3 & *20.5 &\\
& &  &   & 4.28  & 8.0 & 20.1  & \\
BRI$0103+0032$ & 2.87 & 4.383 & 4.437  & 4.23  & 5.8 & 19.8  & \\
BRI$0151-0025$ & 2.74 & 4.142 & 4.194  & &  &     & \\
BRI$0241-0146$ & 2.86 & 4.002 & 4.053  & 3.41 & 5.7 &  19.8  & \\
BR $0245-0608$ & 2.96 & 4.186 & 4.238  & &  &      & \\
BR $0351-1034$ & 3.09 & 4.297 & 4.351  & 3.62 & 6.6 &  19.9  & \\
&  &  &   & 4.14  & 6.3 &   19.9   & \\
BR $0401-1711$ & 2.82 & 4.184 & 4.236 &  &  &     & \\
BR $0951-0450$ & 2.93 & 4.315 & 4.369 & 3.84 & 24.0 & *21.0 & \\
               &      &       &       & 4.20 & 10.6 & *20.3 & \\
BRI$0952-0115$ & 2.99 & 4.372 & 4.426 & 4.01 & 18.0 & *20.8 & \\
BRI$1013+0035$ & 2.61 & 4.351 & 4.405  &3.10 & 17.5 & *20.8  & \\
               &      &       &        &3.73 & 9.6 & 20.2  & \\
               &  &  &   & 4.15 &  7.4  &  20.0   & \\
BR $1033-0327$ & 2.91 & 4.454 & 4.509 &4.15 & 9.6  & 20.2  & \ 20.15 $\pm$ 0.11\\
BRI$1050-0000$ & 2.83 & 4.233 & 4.286  & &  &    & \\
BRI$1108-0747$ & 2.64 & 3.873 & 3.922  & 2.79 & 8.3  &  20.1    & \\
 &  &  &  & 3.607 & 9.0  &  20.2 &*20.33 $\pm$ 0.15\\
BRI$1110+0106$ & 2.58 & 3.869 & 3.918  & 3.25 &  5.2 &  19.7  & \\
& & & &   3.28 & 6.2 & 19.9   & \\
BRI$1114-0822$ & 3.19 & 4.440 & 4.495   &  3.91 & 6.7 & 19.9  & \\
& & & &4.25 & 11.7 & *20.4  & \\
& & & &  4.45 & 5.3 & 19.7  & \\
BR $1144-0723$& \multicolumn{7}{l}{Removed from sample.}  \\
BR $1202-0725$ & 3.16 & 4.637 & 4.694 &  3.20 &  5.4  & 19.7   & \\
& & &  &  3.38 &  7.1  & 20.0   & \\
& & &  &  4.13 &  7.8  & 20.1   & \\
& & &  &4.383 & 13.2 &  20.5  & *20.49 $\pm$ 0.15\\
BRI$1328-0433$ & 2.24 & 4.165 & 4.217  & 3.08 & 8.3 & 20.1  & \\
BRI$1335-0417$ & 3.08 & 4.342 & 4.396   & &   & \\
BRI$1346-0322$ & 2.65 & 3.942 & 3.992  & 3.15 & 6.8 & 19.9  & \\
 &  &  &  &  3.36 & 5.0 &  19.7   & \\
 &  &  &  & 3.73 & 10.0 & *20.3 & \\
BRI$1500+0824$ & 2.39 & 3.894 & 3.943 & 2.80 & 11.3 & *20.4 & \\
GB $1508+5714$ & 2.73 & 4.230 & 4.283   &  &  &   & \\
MG $1557+0313$ & 2.66 & 3.842 & 3.891   &  &  &   & \\
GB $1745+6227$ & 2.47 & 3.852 & 3.901   &  &  &   & \\
BR $2212-1626$ & 2.69 & 3.940 & 3.990   &  &  &   & \\
BR $2237-0607$ & 2.96 & 4.502 & 4.558 & 4.08 & 11.5 & *20.4 & \\
BR $2248-1242$ & 2.94 & 4.109 & 4.161  & &  &   & \\
\hline
\end{tabular}
\raggedright \\
{* These absorbers are above the statistical sample
threshold of \nhi\ $\ge 2 \times 10^{20}$ atoms cm$^{-2}$. } \\
{z$_{min} =$ minimum redshift at which a DLA could be observed}  \\
{z$_{em}  =$ emission redshift of the QSO} \\
{z$_{max} =$ 3000 km s$^{-1}$ blueward of z$_{em}$}\\
{z$_{dla} =$ redshift at which a damped candidate was observed} \\
\end{table}
\twocolumn

\section{The HI Column Density Distribution for log N$_{\rm HI}$ $>$ 17.2}

\subsection{Background}

The distribution of the \hi\ column densities for QSO absorption
line systems has been investigated by several authors.  
Tytler (1987) found that the distribution may be represented
by a single power law $f(N) \propto N^{-\beta}$  over the range 
$13.3 \le$ log \nhi $ \le 21.8$ with $\beta=1.51\pm0.02$.  
He argued that this was evidence for a single population.
Fitting the higher and lower column density systems separately,
Tytler found $\beta=1.61\pm0.10$ for $\log$ \nhi $ < 16.7$ and
$\beta=1.35\pm0.07$ for $\log$ \nhi $ > 17.2$, very similar slopes
particularly considering the inadequacies of the data set. 
Sargent \etal\ (1980) argued that the forest lines and higher column
density metal absorbers were distinct populations due to their
different clustering properties. Sargent, Steidel \& Boksenberg (1989) found a 
single power law fit the over the entire column density range
but noted that since little detailed data was available for 
$15.5 \le$ log \nhi $ \le 17.2$, the region where the \hi\ becomes 
optically thick,
that this was not evidence for a single population.  They 
also found $\beta=1.39$ for $\log$ \nhi $ \ge 17.3$.
LWTLMH found the best fit for the high column
density data ($17.2 \le$ log \nhi $\le 21.8$) gave $\beta=1.25$.
All of these values for $\beta$ agree within the errors.
Petitjean \etal\ (1993) reviewed these results and concluded
that neither a single nor double power fit well 
for $13.3 \le$ \nhi $\le 21.8$ but that there clearly was 
a flattening of the distribution function around $\log$ \nhi $ \approx 16$.

Looking at the damped \lya systems alone ($20.3 \le$ log \nhi $\le 21.8$)
LWTLMH found $\beta=1.73$. 
Most of these investigators find no evidence for any substantial
redshift evolution in the distribution function 
though LWT find an increased number of the
highest column density systems at high redshift.
Into this quagmire we wade with the absorption systems from 
the APM Damped \lya Survey [Paper II] and the results
for the Lyman-limit system evolution from Paper I to
attempt to better quantify what is happening with the high column 
density \hi\ at high redshift.  

\subsection{A Power Law Distribution Function}

The form of the power law column density distribution 
function $f(N)$ typically used is 
\begin{equation}
f(N) = kN^{-\beta}.
\label{fneqn}
\end{equation}
$f(N)dNdX$ is defined as the number of absorbers in 
an absorption distance interval $dX$ with \hi\ column density
$N$ in the range $N+dN$.  The absorption distance $X$ is 
used to remove the redshift dependence in the sample
and put everything on a comoving coordinate scale since $\Delta z=0.5$ 
at redshift 2 is not the same at as a $\Delta z=0.5$
at redshift 4.  
If the population of absorbers is nonevolving (\ie\ their
number density multiplied by their cross-section does not 
change with redshift), the absorption distance can be defined as  
\begin{equation}
X(z) = \int_0^z = (1+z)(1+2q_0z)^{-1/2}dz,
\label{absdiseqn}
\end{equation}
therefore
\begin{equation}
X(z) = \cases {{2 \over 3}[(1+z)^{3/2} - 1] &if q$_0=0.5$;\cr
              {1 \over 2}[(1+z)^2 - 1]    &if q$_0=0$.}
\label{absdis2eqn}
\end{equation}
(Bahcall \& Peebles 1969; \cf\ Tytler 1987).  
The value of $q_0$ has little effect on the slope of the
column density distribution function.
In the analysis below
we have utilised $q_0=0.5$.
To allow for redshift evolution, the distribution function is normally
generalised as  
\begin{equation}
f(N,z) = kN^{-\beta}(1+z)^{\gamma}.
\label{fnzeqn}
\end{equation}
The damped \lya systems and candidates from the APM survey 
shown in table~\ref{t_dla} have been combined with
previous surveys for damped \lya systems (WTSC, LWTLMH, LWT) which
results in a data set of 366 QSOs yielding 44 damped systems
with log \nhi $\ge 20.3$ covering the redshift range $0.2 \le z \le 4.4$.
The total redshift path and absorption distance covered
by the surveys is shown in table~\ref{t_zpath}.
A maximum likelihood technique, described in Appendix A,  
has been employed 
to find values for $\beta$ and $k$ to determine if a power
law fit describes the \hi\ column density distribution for this 
sample.  
As already indicated a disadvantage of a power
law model for the HI column density distribution of damped Ly$\alpha$ 
absorbers, is the divergent nature of the integral mass contained in the
systems.  Since it is straightforward to generalise the maximum likelihood
method to alternative forms of the distribution, we explore in section 3.6
an alternative parameterisation based on a gamma distribution 
(\cf\ Pei \& Fall 1995). 

\begin{table}
\caption{Redshift and Absorption Distance Paths}
\label{t_zpath}
\begin{tabular}{lrrr}
\hline
Data Set & $\Delta z$ &\multicolumn{1}{c}{$\Delta X$}& \multicolumn{1}{c}{$\Delta X$} \\
 & & ($q_0=0$) & ($q_0=0.5$)  \\
\hline
{\bf $0 < z < 4.7$} & & & \\
APM Damped \lya Survey  &  36.1 & 162.3 &  76.5  \\
WTSC $+$ LWTLMH $+$ LWT & 203.4 & 602.1 & 344.8  \\
Combined                & 239.5 & 764.4 & 421.3  \\
{\bf $z > 3$} & & & \\
APM Damped \lya Survey  &  30.5 & 141.2 &  65.5  \\
WTSC $+$ LWTLMH $+$ LWT &  13.0 &  54.6 &  26.7  \\
Combined                &  43.5 & 195.8 &  92.2  \\
 & & & \\
\hline
\end{tabular}
\end{table}

\subsection{Results for a Single Power Law}

A single power law fit to the combined data set described
above of the form in equation~\ref{fnzeqn}
with $q_0=0.5$ results in $\beta=1.74\pm0.12$ and 
$\log k=13.9\pm2.3$ with similar values for $q_0=0$ 
($\beta=1.75, \log k=13.9$). The quoted error for
the normalisation constant $k$ is large because for
a small change in the value of $\beta$, $k$ can change
by 2 orders of magnitude. These results are in
good agreement with the results
found by LWTLMH ($\beta=1.73\pm0.29$, log k$=13.63\pm0.09$)
and are plotted for the entire data set in  
figure~\ref{f_dlafn_all}. The data is binned for 
display purposes only with the vertical error bars 
plotted at the mean column density for each bin.  
We will see in the next section that 
a single power law is not a good fit to the data. 

\subsection{Cumulative HI Distribution}

As shown in Paper I for the Lyman-limit system evolution,
the arbitrary binning of the data for presentation in differential plots
includes a subjective component that can mask exactly what is 
happening in the underlying data.
A cumulative distribution plot  
is far better at revealing the true nature of the distribution and 
this approach is examined now. 
The $\log$ of the cumulative number of damped \lya systems detected  
versus $\log$ \nhi\ is plotted in figure~\ref{f_dlacum}(a).  A point
for the expected number of Lyman-limit systems that
would be detected down to $\log$ \nhi $=17.2$ is shown with a circled star.
This is calculated by integrating the number density per unit 
redshift ($N(z)=0.27(1+z)^{1.55}$) over the redshift path covered by 
the $n$ QSOs in the DLA sample, \ie\
\begin{eqnarray*}
{\rm LLS}_{expected} &=& \sum_{i=1}^n \int N(z)dz = 
	 \sum_{i=1}^n \int_{z_{min}}^{z_{em}} N_0(1+z)^{\gamma}dz \\
  &=&  \sum_{i=1}^n \int_{z_{min}}^{z_{em}} 0.27(1+z)^{1.55}dz.
\end{eqnarray*}
\begin{equation}
\label{expllseqn}
\end{equation}
It is obvious from figure 3(a) that a power law will not 
fit the entire column density range 17.2 $\le$ log \nhi $\le$ 22.
A Kolmogorov-Smirnov (K-S) yields a probability of less than $10^{-7}$
that the fit represents the underlying data set.   
In figures~\ref{f_dlacum}(b-d) the same distribution is 
overplotted with single power law fits for different 
values of $\beta$ that were fit to the graph by eye.
(b) shows that $\beta=1.34$ will fit from the Lyman limit column 
density through the damped systems with $\log$ \nhi $\approx21$,
a flatter slope than the canonical 1.5-1.7 range.  
(c) shows that $\beta=1.69$ fits the damped distribution 
with $20.3\le\log$ \nhi $\le21.2$ well but does not describe the 
high or low column density tails of the distribution.
(d) shows a fit to the sharp drop off in numbers for damped systems
with $\log$ \nhi \simgt\ 21.3.  This can be expected from looking 
at the estimated column densities for the damped systems 
in table~\ref{t_dla} or by looking at the spectra.  There 
are not a lot of heavily damped systems. 
Clearly the results for a single power law fit depend critically 
on the range of column densities included.  This characteristic 
can explain much of 
the variation in the results previously 
seen by various authors that were summarised
in section 3.1. 

\subsection{Redshift Evolution for 0.008 $<$ z $<$ 4.7}

To qualitatively study the redshift evolution of the 
column density distribution of the damped \lya systems
the cumulative distribution shown in figure~\ref{f_dlacum} 
has been split in half at redshift 2.5 and each set plotted
individually (figure~\ref{f_zsplit}).  
The damped \lya systems with $z>2.5$ 
are shown by the solid line and the absorbers with $z<2.5$ are shown 
by the dashed line.  The higher redshift absorbers appear to have 
a slightly flatter slope up to $\log$ \nhi $=$21 and then a sharper
drop in the number of very high column density systems,  
though a K-S test shows that this is not a statistically significant
difference.  
The evolution with redshift in the slope of the column density
distribution is also apparent when looking at the differential
$f(N)$.  LWT plotted this in 3 redshift bins z$=$[0.008,1.5],[1.5,2.5],
and [2.5,3.5].  In the highest redshift bin there was a flattening
of the column density distribution slope towards higher column 
densities.  In figure~\ref{f_dlafn} we have plotted our combined 
data set with this same
binning with the addition of one higher redshift bin z$=$[3.5,4.7]. 
The flattening of the distribution function
towards higher column density systems in the z$=$[2.5,3.5] bin
in the LWT data is no longer pronounced.  The most striking feature is the  
steepness of the distribution in the highest redshift bin.   
It is not just steeper due to a decrease in the highest column 
density systems (log \nhi$>$21), but there is also an increase
in the number of lower column density systems relative to the other bins.
Even if 15-20\% of the candidate systems with log \nhi$\approx$20.3
turn out not to be damped when observed at higher resolution, as we 
expect, this result still holds. 

\subsection{Results for a Gamma-Distribution}

There are two strong motivating factors to find an alternative
model for describing the HI column density distribution.  First, as shown
in section 3.4, there is direct evidence for an apparent variation in the
power law slope as a function of $N_{\rm HI}$.  This implies that a higher
order functional form other than a power law is needed to describe 
the column
density distribution. Second, as noted in the introduction, with a power law
model the integral mass contained within damped Ly$\alpha$ systems is divergent
for realistic values of $\beta$.  This in turn means that it is impossible
to assign a formal upper limit to any estimate of the neutral gas content of
the early Universe.  Consequently, following Pei and Fall (1995), we have
chosen to model the data with a gamma distribution of the form
\begin{equation}
f(N,z)=( f_* / N_* ) ( N / N_* )^{-\beta} e^{-N/N_*}
\label{gamdisteqn}
\end{equation}
where $f_*$ is the characteristic number of absorbing systems at the column
density $N_*$, and $N_*$ is a parameter defining the turnover, or `knee',
in the number distribution.  Both $f_*$ and $N_*$ may in general vary with
redshift but for the moment we treat them as constants.  This functional
form is similar to the Schechter luminosity function (Schechter 1976).
For $N << N_*$ the gamma function tends to the same form as the single power
law, $f(N) \propto N^{-\beta}$; whilst for $N$ \simgt\ $N_*$, the exponential term
begins to dominate.

We can understand how a gamma function might provide a better
description of the damped Ly$\alpha$ data by considering the differential
logarithmic slope which is given by
\begin{equation}
{d log f(N,z) \over d log N} = - \beta - {N \over N_*}
\label{gamdiffslopeeqn}
\end{equation}
As the column density approaches $N_*$ the slope begins to steepen and rapidly
turns over at higher column densities, qualitatively similar to what we
observe in figures~\ref{f_dlacum} and~\ref{f_zsplit}.
Furthermore, the integral HI over the column
density distribution (\cf\ equation~\ref{himass1}) 
for $N_{min} << N_*$ is now given by
$f_* N_* \tau (2 - \gamma)$ where $\tau$ denotes the standard gamma function.
This function is bounded if $\gamma < 2$.

The maximum likelihood technique outlined in Appendix A can readily be modified
to incorporate this form.  We note that the likelihood solution can 
be found over a two-dimensional grid of pairs of values of $N_*$ and
$\beta$, since the constant $f_*$ can be directly computed 
using the constraint
\begin{equation}
m =\sum_{i=1}^n f_* \int_{N_{min}}^{N_{max}} \int_{z_{min}^i}^{z_{max}^i} f(N,z) dz dN
\label{gamconstrainteqn}
\end{equation}
where m is the total number of observed systems. 
This is computationally much 
less intensive than doing a 3-D grid search. The results of a single
functional fit to the entire dataset are log $N_*=21.63\pm0.35$,
$\beta=1.48\pm0.30$, and $f_*=1.77\times 10^{-2}$.
The log-likelihood function results with confidence 
contours are shown in figure~\ref{f_gamconf}(a).  The best fit 
is overplotted on the differential form of $f(N)$ in 
figure~\ref{f_gamconf}(b) and on the cumulative distribution
in figure~\ref{f_gamconf}(c). (The single power law form of 
$f(N)$ was shown fitted to the same data in 
figures~\ref{f_dlafn_all} and~\ref{f_dlacum}).  
The differential form of the plots (figures~\ref{f_dlafn_all}
and~\ref{f_gamconf}(b)) 
show little difference between the single power 
law and $\Gamma$-distribution fits.  When displayed
with the cumulative number of absorbers in figure~\ref{f_gamconf}(c),
the $\Gamma$-distribution now clearly fits the entire 
data set with column densities log \nhi $\ge$ 20.3. 
If the expected number of Lyman-limit systems are included
in the fit, the results are log $N_*=21.36\pm0.15$,
$\beta=1.16\pm0.15$, and $f_*=4.43\times 10^{-2}$. This 
also provides a reasonable fit to the data as shown in 
figure~\ref{f_gamconf2}(a-c).

\section{Number Density Evolution with Redshift}

Differential evolution in the number density
of damped \lya absorbers has been described by LWT and
Wolfe \etal\ (1995).
While the change in number density per unit redshift is 
consistent with no intrinsic evolution of the absorbers
over the range 0 $<$ z $<$ 3.5,
they find that the systems with log \nhi$>21$ disappear
at a much faster rate from z$=$3.5 to z$=$0 than does
the population of damped absorbers as a whole.
We now examine the redshift
evolution of the damped \lya absorbers in our combined data set
by determining the  number density of absorbers
per unit redshift, $dN/dz \equiv N(z)$.  In a standard Friedmann Universe for
absorbers with cross section $\pi{R_{0}}^2$ and number density  $\Phi_{0}$
per unit comoving volume
\begin{equation}
N(z) = \Phi_{0}\pi{R_{0}}^2cH_{0}^{-1}(1 + z)(1 + 2q_{0}z)^{-1/2}.
\label{sybteqn}
\end{equation}
It is customary to represent the number density as a power law of the form
\begin{equation}
N(z) = N_{0}(1 + z)^\gamma,
\label{nzeqn}
\end{equation}
where $N_{0} = \Phi_{0}\pi{R_{0}}^2cH_{0}^{-1}$.
This yields $\gamma$ = 1 for $q_{0}$ = 0 and
$\gamma$ = 1/2 for $q_{0}$ = 1/2 for the case of no evolution with redshift
in the product of the number density and cross section of the absorbers
\cite{SYBT}.

A maximum likelihood fit to the data yields
$N(z)=0.04(1+z)^{1.3\pm0.5}$ which is consistent with
no intrinsic evolution
even though the value of $\gamma$ is similar
to that found for the Lyman limit
systems where evolution is detected at a significant level
(Paper I; Stengler-Larrea \etal\ 1995).
The log-likelihood function for $\gamma$ and $N_0$ with
$>$68.3\% and  $>$95.5\% confidence contours
is plotted in figure~\ref{f_maxz}.
We also find redshift evolution in the higher
column density systems but with a decline in $N(z)$
for z$>$3.5.
These results are displayed in figure~\ref{f_dndz}.
The entire data set is plotted as dashed lines with the
above fit. The results for only the absorbers with
log N(HI)$\ge21$ are shown as solid lines.
Figure~\ref{f_zhi} shows HI column density versus redshift,
and the paucity of absorbers with log \nhi $>$ 21 at z $>$ 4 
is apparent.

\section{Conclusions}
Three QSOs from APM survey have been observed at 0.8\AA\ resolution.
Two have damped systems with confirmed HI column densities
of \nhi $\ge$ 10$^{20.3}$ atoms cm$^{-2}$, with a third
absorber falling just below this threshold.
We have discovered the highest redshift damped \lya 
absorber known at z$=$4.383 in QSO BR1202$-$0725. 
The two systems with \nhi $\ge$ 10$^{20.3}$ atoms cm$^{-2}$, 
and remaining nine 
candidate damped absorbers from the APM survey have 
been combined with data from previous surveys 
to study the column density distribution and 
number density evolution for absorbers with \nhi\ $\ge$ 17.2.
If the HI column density distribution function is fit
with a power law, $f(N)=kN^{-\beta}$, we find evidence 
for breaks in the power law, flattening for 
17.2 $\le$ log \nhi\ \simlt\ 21 and steepening for
log \nhi\ \simgt\ 21.2. 
The column density distribution function
for the data with log\nhi$\ge$20.3 is
better fit with the $\Gamma$-distribution form 
$f(N)=(f_*/N_*)(N/N_*)^{-\beta}exp(-N/N_*)$ with 
log $N_*=21.63\pm0.35$,
$\beta=1.48\pm0.30$, and $f_*=1.77\times 10^{-2}$.
 
For the number density evolution of the damped 
absorbers (log \nhi\ $\ge$ 20.3) over the redshift 
range 0.008 $<$ z $<$ 4.7 we find the best
fit of a single power law form for  
$N(z)=N_0(1+z)^{\gamma}$ yields 
$\gamma=1.3\pm0.5$ and $N_0=.04^{+.03}_{-.02}$.
This is consistent with no intrinsic evolution
in the absorbers even though the value of $\gamma$ is similar
to that found for the Lyman limit
systems where evolution is detected at a significant level.
Evolution is evident in the highest
column density absorbers with the incidence of
systems with log N(HI)$\ge$21
decreasing for z \simgt\ 3.5.

\noindent{\bf Acknowledgments}

We would like to thank Bob Carswell for providing software for
and assistance with the data reduction and profile fitting of the spectra.
LSL acknowledges support from an Isaac Newton Studentship, the Cambridge
Overseas Trust, and a University of California President's Postdoctoral
Fellowship.  RGM acknowledges the support of the Royal Society.

\begin{appendix}
\section{Maximum Likelihood Analysis}

Using equation~\ref{fnzeqn} for the column density 
distribution function, the damped \lya absorbers will be found 
randomly distributed according to this function along the 
QSO line-of-sight in $N-z$ space.
If the space is divided into $m$ cells each of volume $\delta v$,
the expected number of points in cell $i$ is given by
\begin{equation}
\phi_i = f(N,z)_i \delta v.
\label{fnexpeqn}
\end{equation}
The probability of observing $x_i$ points in cell $i$ is
\begin{equation}
p(x_i) = e^{-\phi_i} {\phi_i^{x_i} \over x_i!}.
\label{pxieqn}
\end{equation}
The likelihood function for QSO$_j$ taking the product over
all the cells is then
\begin{equation}
L_j = \prod_{i=1}^m p(x_i) = \prod_{i=1}^m  
e^{-\phi_i} {\phi_i^{x_i} \over x_i!}.
\label{dlal1eqn}
\end{equation}
If the volume of each cell $\delta v$ becomes very small such 
that there is either 1 or 0 points in each cell, 
$${x_i} = \cases {1,&if DLA detected;\cr 0,&if none detected,}$$
then the likelihood can be rewritten separating out the terms for
full and empty cells.  For $m = g$ empty cells $+$ $p$ full cells
\begin{equation}
L_j = \prod_{i=1}^g e^{-\phi_i}  \prod_{j=1}^p  
e^{-\phi_j} \phi_j
= \prod_{i=1}^m e^{-\phi_i}  \prod_{j=1}^p  \phi_j
\label{dlal2eqn}
\end{equation}
Taking the $\log$ of the likelihood function gives
\begin{eqnarray*}
\log L_j &=& \sum_{i=1}^m -\phi_i +  \sum_{j=1}^p  \ln \phi_j \\
         &=& \sum_{i=1}^m -f(N,z)_i \delta v + \sum_{j=1}^p \ln f(N,z)_j 
	 + p\ln\delta v 
\end{eqnarray*}
\begin{equation}
\label{dlal3eqn}
\end{equation}
(\cf\ Schechter \& Press 1976).
Ignoring the constant terms, in the limit where 
$\delta v \rightarrow 0$ this becomes
\begin{eqnarray*}
\log L_j &=& - \int_{z_{min}}^{z_{max}}  \int_{N_{min}}^{N_{max}} 
             f(N,z)dNdz + \sum_{j=1}^p \ln f(N,z)_j  \\
         &=& - \int_{z_{min}}^{z_{max}}  \int_{N_{min}}^{N_{max}} 
	 kN^{-\beta}(1+z)^{\gamma}dNdz + \sum_{j=1}^p \ln 
	 \bigl[kN^{-\beta}(1+z)^{\gamma}\bigr] 
\end{eqnarray*}
\begin{equation}
\label{dlal4eqn}
\end{equation}
To get the overall log likelihood for $n$ QSOs we evaluate the integrals in 
equation~\ref{dlal4eqn} and additively combine the $\log$ L's resulting in 
\begin{eqnarray*}
\log L &=& \sum_{i=1}^n \Bigl[ {kN_{min}^{1-\beta} \over (1-\beta)(1+\gamma)}
 \Bigl( (1+z_{em}^i)^{1+\gamma}-(1+z_{min}^i)^{1+\gamma}\Bigr)  +
 p\ln k \\
 &+& \sum_{j=1}^{p_i} \Bigl( -\beta\ln N_j + \gamma\ln(1+z_{dla}^j)\Bigr)
 \Bigr] 
\end{eqnarray*}
\begin{equation}
\label{dlal5eqn}
\end{equation}
where $p_i$ is the number of detected DLAs in QSO$_i$ and $N_{min}$ 
is the minimum column density.
\end{appendix}


\onecolumn

\begin{figure}
\flushleft
\caption[]
{The profile fit and $\pm 1 \sigma$ fits to the damped \lya absorbers 
are shown as solid lines. The error arrays are shown as dotted lines.
The components are listed in table~\ref{t_dla_stats}. 
(a) The \hi\ absorption complex in BR1033$-$0327 at z$=$4.15
is a system of at least 5 absorbers with a total
column density of log \nhi$=20.15\pm0.11$ atoms cm$^{-2}$.
(b) The damped \lya absorber in BRI1108$-$0747 at z$=$3.607
with $\log$ \nhi$=20.33\pm0.15$ atoms cm$^{-2}$.
(c) The damped \lya absorber in BR1202$-$0725 at z$=$4.383.  This is
the highest redshift damped \lya absorber known. The central
damped component is shown with log \nhi$=$20.49$\pm$0.15.}
\label{f_dla}
\end{figure}


\begin{figure}
\flushleft
\caption[Single Power Law Fit for \nhi $\ge20.3$-Whole Sample]
{A single power law form of the column density distribution
function, $f(N)=kN^{-\beta}$, fit to the entire damped \lya sample 
from the APM Damped \lya Survey, WTSC, LWTLMH, and LWT. 
The parameters of the fit are $\beta=1.74$ and $\log k=13.9$.  
}
\label{f_dlafn_all}
\end{figure}


\begin{figure}
\flushleft
\caption[Cumulative Distribution for $17.2\le\log$ {\bf\nhi} $\le22$]
{(a) The cumulative distribution for $17.2\le\log$ {\bf\nhi} $\le22$.
The stepped line is the data for all the damped \lya systems 
in the data set.
The circled point is the number of Lyman limit systems
that would be expected given the redshift path covered in the 
damped \lya surveys.  
In (b)-(d) the same distribution is 
overplotted with single power law fits for different 
values of $\beta$ that were fit to the graph by eye.
(b) shows that $\beta=1.34$ will fit from the Lyman limit column 
density through the damped systems with $\log$ \nhi\ $\approx21$,
a flatter slope than the canonical $\beta\approx1.5-1.7$ range.  
(c) shows that $\beta=1.69$ fits the systems 
with $20.3\le\log$ \nhi\ $\le21.2$ well but does not describe the 
high or low column density tails of the distribution.
(d) shows a fit to the sharp drop off in numbers of damped systems
with $\log$ \nhi\ \simgt\ 21.  This is evident from just looking 
at the estimated column densities for the damped systems 
in table~\ref{t_dla} or by looking at the spectra.  There 
are not a lot of heavily damped systems. 
}
\label{f_dlacum}
\end{figure}


\begin{figure}
\flushleft
\caption[Cumulative HI Column Density Distribution - Redshift Split]
{The cumulative HI column density distribution with the sample
split in half at $z=2.5$. The damped \lya systems with $z>2.5$ 
are shown by the solid line and the absorbers with $z<2.5$ are shown 
by the dashed line.  The higher redshift absorbers appear to have 
a slightly flatter slope up to log \nhi$=21$ and then a sharper
drop in the number of very high column density systems,  
though a K-S test shows that this is not a statistically significant
difference.  
}
\label{f_zsplit}
\end{figure}

\begin{figure}
\caption[]
{The log column density distribution function $f(N)$ 
vs.~the log column density \nhi\ plotted over 4 redshift ranges,
z$=$[0.008, 1.5], [1.5, 2.5], [2.5, 3.5], 
and [3.5, 4.7], all binned in the column density ranges
log \nhi$=$[20.3, 20.5], [20.5, 21.0], and [21.0, 21.8]. 
The gradual flattening of the distribution function 
from redshift z$=$0 to z$=$3.5 is evident.
The most striking feature 
is the steepness of the distribution in the highest redshift bin.  
It is not just steeper due to a decrease in the highest column 
density systems (log \nhi$>$21), but there is also an increase
in the number of lower column density systems.
}
\label{f_dlafn}
\end{figure}


\begin{figure}
\flushleft
\caption[]
{(a)The log-likelihood function for the $\Gamma$-distribution
form of the column density distribution function,
$f(N,z)=(f_*/N_*)(N/N_*)^{-\beta} exp(-N/N_*)$.  
The $>$68.3\%, $>$95.5\%, and $>$99.7\% confidence contours
are plotted for log $N_*$ and $\beta$. The best fit 
values are log $N_*=21.63$, $\beta=1.48$, and 
$f_*=1.77\times10^{-2}$ which is solved for analytically.
(b) The $\Gamma$-distribution
function, $f(N,z)=(f_*/N_*)(N/N_*)^{-\beta} exp(-N/N_*)$,
overplotted on the differential
form of the column density distribution.
The fit is to the entire data set including the surveys
from Paper II, WTSC, LWTLMH, and LWT.  
The parameters for the fit are
log $N_*=21.63$, $\beta=1.48$, and $f_*=1.77\times10^{-2}$.
(c) The cumulative distribution for $17.2\le\log$ {\bf\nhi} $\le22$
as shown in figure~\ref{f_dlacum}(a) is now overplotted with
the $\Gamma$-distribution form of the column density
distribution function. The dashed line now clearly fits
the entire distribution for log \nhi$\ge$20.3.
The circled point is again the number of Lyman limit systems
that would be expected given the redshift path covered in the
damped \lya surveys. This is not included in the fit. 
}
\label{f_gamconf}
\end{figure}


\begin{figure}
\caption[]
{This figure shows the same data plotted in 
figure~\ref{f_gamconf} but the fit includes the expected 
number of Lyman limit systems, given the redshift path 
surveyed.
(a) The log-likelihood function for the $\Gamma$-distribution
form of the column density distribution function,
$f(N,z)=(f_*/N_*)(N/N_*)^{-\beta} exp(-N/N_*)$.  
The $>$68.3\%, $>$95.5\%, and $>$99.7\% confidence contours
are plotted for log $N_*$ and $\beta$. The best fit 
values are log $N_*=21.36$, $\beta=1.16$, and 
$f_*=4.43\times10^{-2}$ which is solved for analytically.
(b) The $\Gamma$-distribution
function, $f(N,z)=(f_*/N_*)(N/N_*)^{-\beta} exp(-N/N_*)$,
overplotted on the differential
form of the column density distribution.
The fit is to the entire data set including the surveys from Paper II,
WTSC, LWTLMH, and LWT.  
The parameters for the fit are
log $N_*=21.36$, $\beta=1.16$, and $f_*=4.43\times10^{-2}$.
(c) The cumulative distribution for $17.2\le\log$ {\bf\nhi} $\le22$
overplotted with
the $\Gamma$-distribution form of the column density
distribution function. 
The circled point is again the number of Lyman limit systems
that would be expected given the redshift path covered in the
damped \lya surveys. 
}
\label{f_gamconf2}
\end{figure}


\begin{figure}
\flushleft
\caption[]
{The $>$68.3\% and $>$95.5\% confidence contours for the
log-likelihood function are plotted for the number density 
per unit redshift of the damped absorbers.
The best fit for a single power law form 
$N(z)=N_0(1+z)^{\gamma}$ yields
$\gamma=1.3\pm0.5$ and $N_0=.04^{+.03}_{-.02}$ 
over the redshift range 0.008 $<$ z $<$ 4.7.}
\label{f_maxz}
\end{figure}


\begin{figure}
\flushleft
\caption[]
{The number density of DLA per unit redshift,
$N(z)$, vs. z(absorption). The dashed bins show
N(z) for all the damped systems and the solid bins
for systems with N(HI)$\ge10^{21}$ atoms cm$^{-2}$.
A single power law fit of N(z)$=.04(1+z)^{1.3}$ 
is overplotted. This is 
consistent with no intrinsic evolution
in the absorbers even though the value of $\gamma$ is similar
to that found for the Lyman limit
systems where evolution is detected at a significant level
(Paper I).}
\label{f_dndz}
\end{figure}

\begin{figure}
\flushleft
\caption[]
{The HI column density of the damped \lya absorbers
is plotted versus absorption redshift.  The paucity 
of absorbers with log \nhi $>$ 21 at z $>$ 4 is apparent.}
\label{f_zhi}
\end{figure}


\begin{thebibliography}{}

\bibitem[\protect\citename{{Bahcall} \& {Peebles}}{1969}]{BP69}
{Bahcall} J.N., {Peebles} P.J.E., 1969, ApJ,{ 156}, L7

\bibitem[\protect\citename{{Carswell} {et~al.} }{1984}]{Carswell84}
{Carswell} R.F., {Morton} D.C., {Smith} M.G., {Stockton} A.N., {Turnshek} D.A.,
  {Weymann} R.J., 1984, ApJ,{ 278}, 486

\bibitem[\protect\citename{{Cooke}}{1994}]{Cooke94}
{Cooke} A.J., 1994, Ph.D. Thesis.
\newblock Cambridge University

\bibitem[\protect\citename{{Fall} et~al. }{1989}]{FPM89}
{Fall} S.M., {Pei} Y.C., {McMahon} R.G., 1989, ApJ,{ 341}, L5

\bibitem[\protect\citename{{Giallongo} et~al. }{1994}]{Giallongo94}
{Giallongo} E., {D'Odorico} S., {Fontana} A., {McMahon} R.G., {Savaglio} S.,
  {Cristiani} S., {Molaro} P., {Trevese} D., 1994, ApJ,{ 425}, L1

\bibitem[\protect\citename{{Lanzetta} et~al. }{1995}]{LWT95}
{Lanzetta} K.M., {Wolfe} A.M., {Turnshek} D.A., 1995, ApJ,{ 440}, 435

\bibitem[\protect\citename{{Lanzetta} et~al. }{1991}]{LWTLMH91}
{Lanzetta} K.M., {Wolfe} A.M., {Turnshek} D.A., {Lu} L., {McMahon} R.G.,
  {Hazard} C., 1991, ApJS,{ 77}, 1

\bibitem[\protect\citename{{Lu} et~al. }{1996}]{Lu96}
Lu L, Sargent W.L.W., Womble D.S., Barlow T.A., 1996, ApJ, 457, L1

\bibitem[\protect\citename{{Pei}, {Fall} \& {Bechtold}}{1991}]{PFB91}
{Pei} Y.C., {Fall} S.M., {Bechtold} J., 1991, ApJ,{378}, 6

\bibitem[\protect\citename{{Pei} \& {Fall} }{1995}]{PF95}
{Pei} Y.C., {Fall} S.M., 1995, ApJ,{ 454}, 69

\bibitem[\protect\citename{{Petitjean} et~al. }{1993}]{Petitjean93}
{Petitjean} P., {Webb} J.K., {Rauch} M., {Carswell} R.F., {Lanzetta} K.M.,
  1993, MNRAS,{ 262}, 499

\bibitem[\protect\citename{{Pettini} et~al. }{1990}]{PBH90}
{Pettini} M., {Boksenberg} A., {Hunstead} R.W., 1990, ApJ,{ 348}, 48

\bibitem[\protect\citename{{Pettini} et~al. }{1994}]{PSHK94}
{Pettini} M., {Smith} L.J., {Hunstead} R.W., {King} D.L., 1994, ApJ,{ 426}, 79

\bibitem[\protect\citename{{Rauch} et~al. }{1990}]{Rauch90}
{Rauch} M., {Carswell} R.F., {Robertson} J.G., {Shaver} P.A., {Webb} J.K.,
  1990, MNRAS,{ 242}, 698

\bibitem[\protect\citename{{Sargent} et~al. }{1989}]{SSB}
{Sargent} W.L.W., {Steidel} C.C., {Boksenberg} A., 1989, ApJS,{ 79}, 703

\bibitem[\protect\citename{{Sargent} et~al. }{1980}]{SYBT}
{Sargent} W.L.W., {Young} P.T., {Boksenberg} A., {Tytler} D., 1980, ApJS,{ 42},
  41

\bibitem[\protect\citename{{Schechter }}{1976}]{Schechter76}
{Schechter} P., 1976, ApJ,{ 203}, 297

\bibitem[\protect\citename{{Schechter} \& {Press}}{1976}]{SP76}
{Schechter} P., {Press} W.H., 1976, ApJ,{ 203}, 557

\bibitem[\protect\citename{{Stengler-Larrea} et~al. }{1995}]{Stengler95}
Stengler-Larrea E.A., {\etal},
1995, ApJ, 444, 64

\bibitem[\protect\citename{{Storrie-Lombardi} {et~al.} }{1994}]{SMIH94}
{Storrie-Lombardi} L.J., {McMahon} R.G., {Irwin} M.J., {Hazard} C., 1994, ApJ,{
  427}, L13 (Paper I)

\bibitem[\protect\citename{{Storrie-Lombardi} et~al. }{1996}]{SMIH96}
{Storrie-Lombardi} L.J., {McMahon} R.G., {Irwin} M.J., {Hazard} C., 1996, ApJ, 468, 121 (Paper II) 

\bibitem[\protect\citename{{Storrie-Lombardi}, {McMahon}, \& {Irwin} }{1996}]{SMI96}
{Storrie-Lombardi} L.J., {McMahon} R.G., {Irwin} M.J., 
1996, MNRAS, in press (Paper IV)

\bibitem[\protect\citename{{Tytler}}{1987}]{Tytler87}
{Tytler} D., 1987, ApJ,{ 321}, 49

\bibitem[\protect\citename{{Wampler} et~al. }{1996}]{Wampler96}
{Wampler} E.J., {Williger} G.M., {Baldwin} J.A., {Carswell} R.F., {Hazard} C.,
  {McMahon} R.G., 1996, A\&A, in press 

\bibitem[\protect\citename{{Williger} et~al. }{1994}]{Williger94}
{Williger} G.M., {Baldwin} J.A., {Carswell} R.F., {Cooke} A.J., {Hazard} C.,
  {Irwin} M.J., {McMahon} R.G., {Storrie-Lombardi} L.J., 1994, ApJ,{ 428}, 574

\bibitem[\protect\citename{{Wolfe}}{1987}]{Wolfe87}
{Wolfe} A.M., 1987, Proc.Phil.Trans.Roy.Soc.,{ 320}, 503

\bibitem[\protect\citename{{Wolfe} et~al. }{1986}]{Wolfe86}
{Wolfe} A.M., {Turnshek} D.A., {Smith} H.E., {Cohen} R.D., 1986, ApJS,{ 61},
  249

\bibitem[\protect\citename{{Wolfe} et~al. }{1995}]{WLFC95}
{Wolfe} A.M., {Lanzetta} K.M., Foltz C.B., Chaffee F.H., 1995, ApJ, 454, 698

\end{thebibliography}
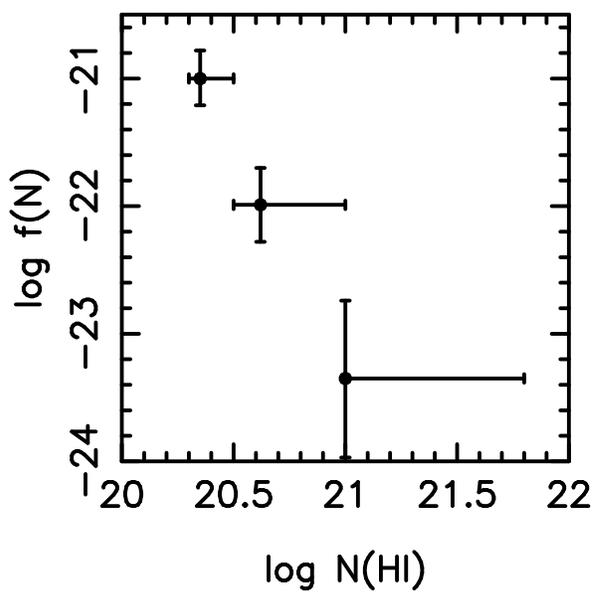
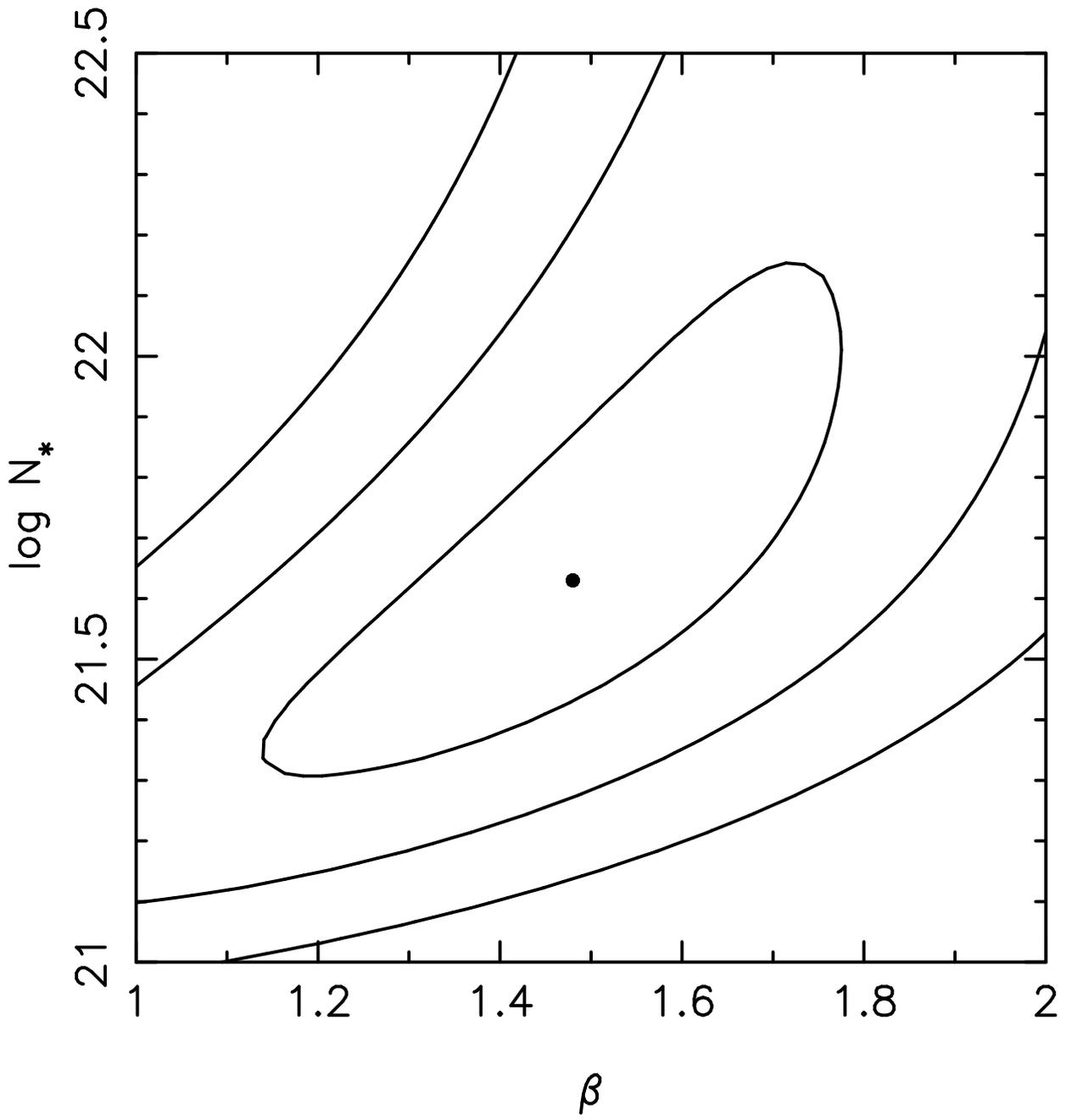
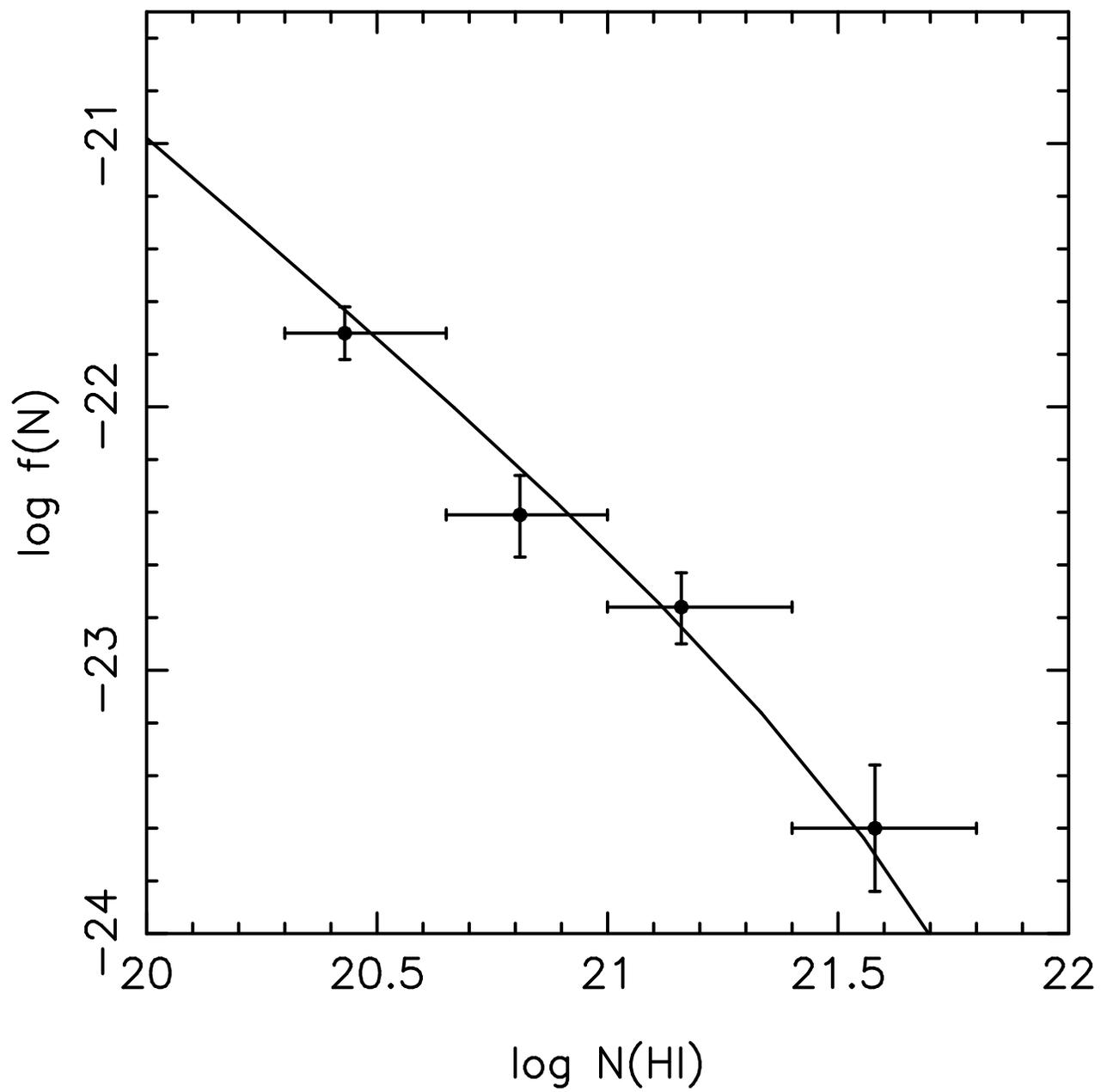
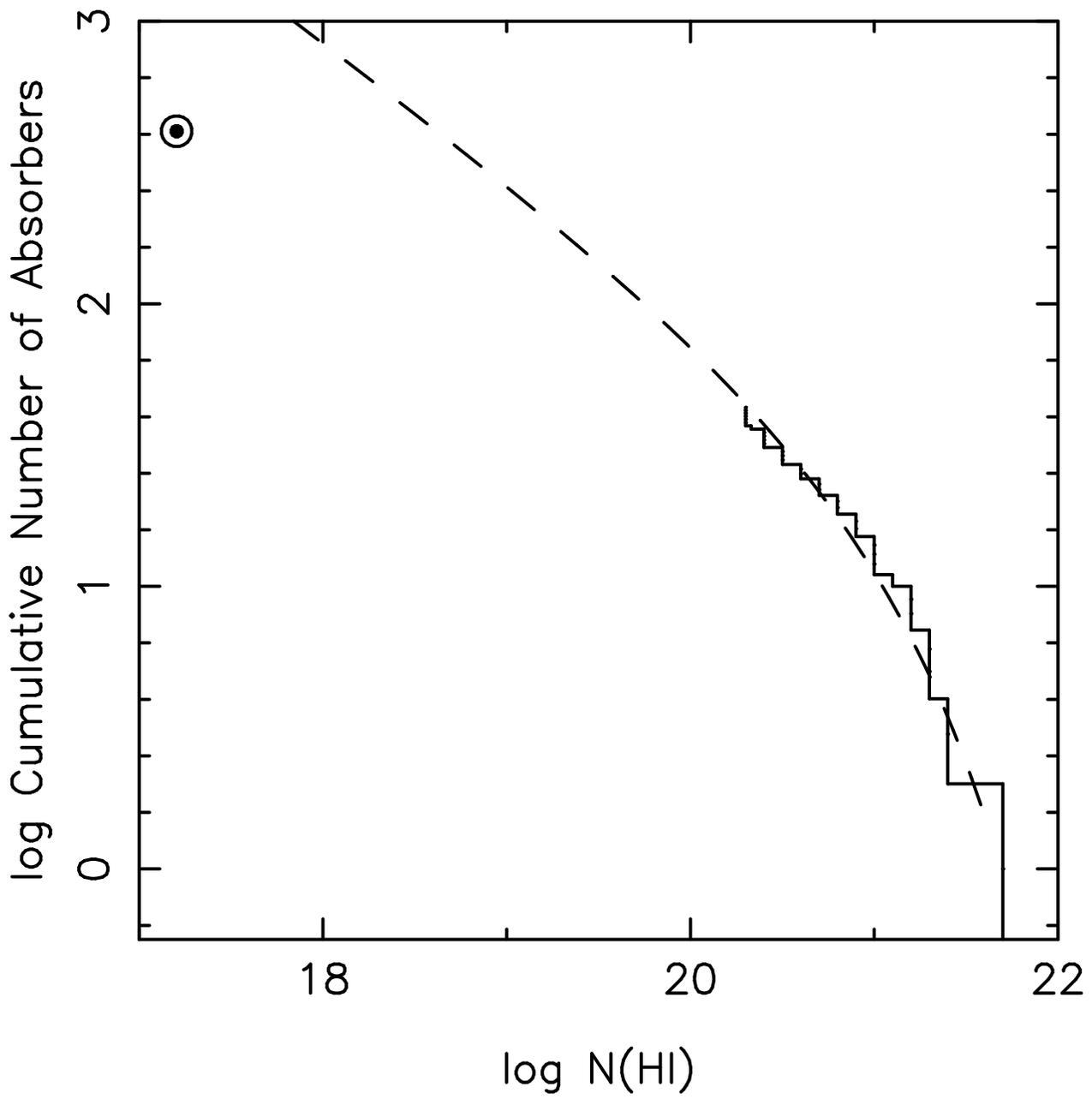
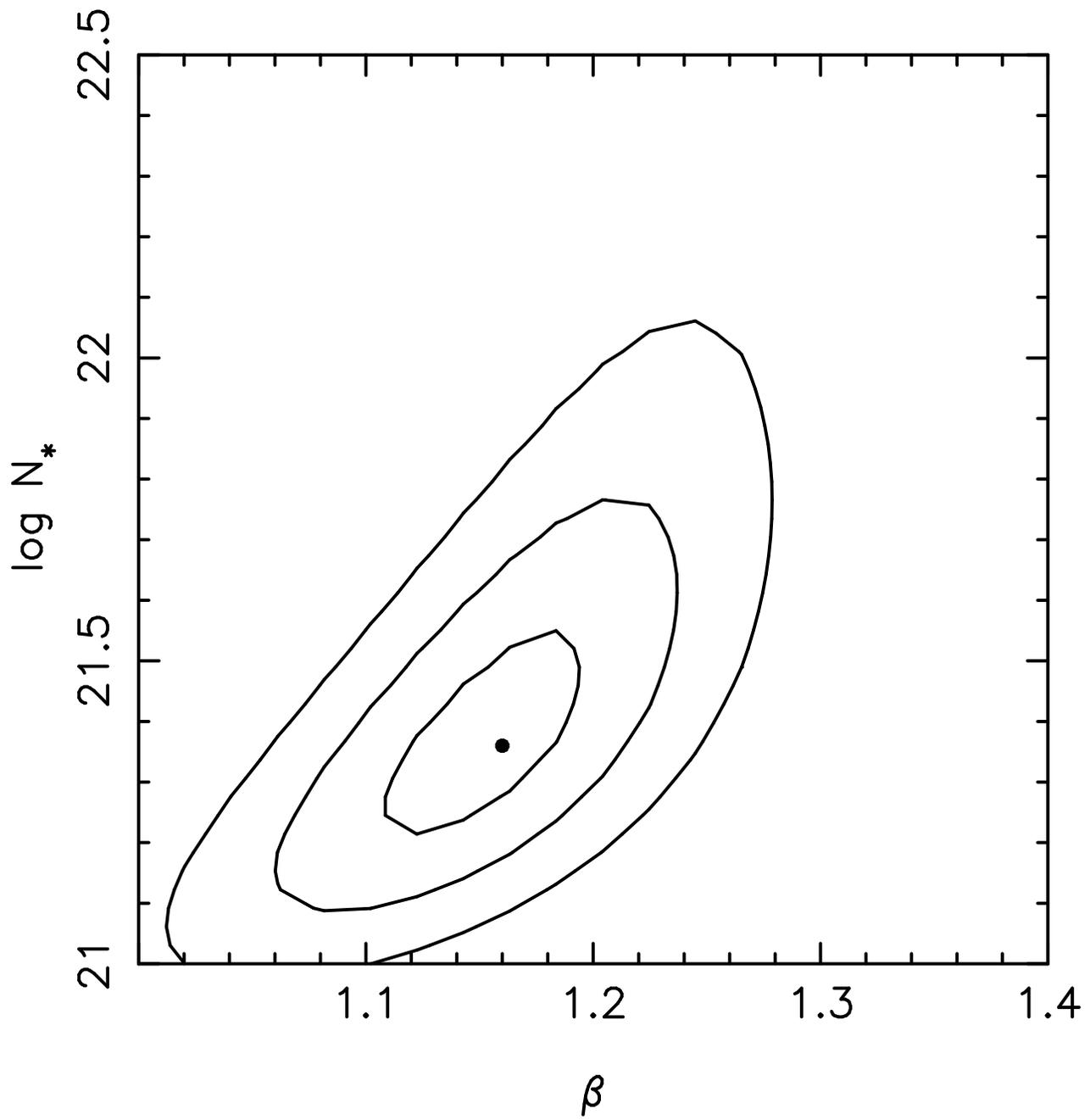
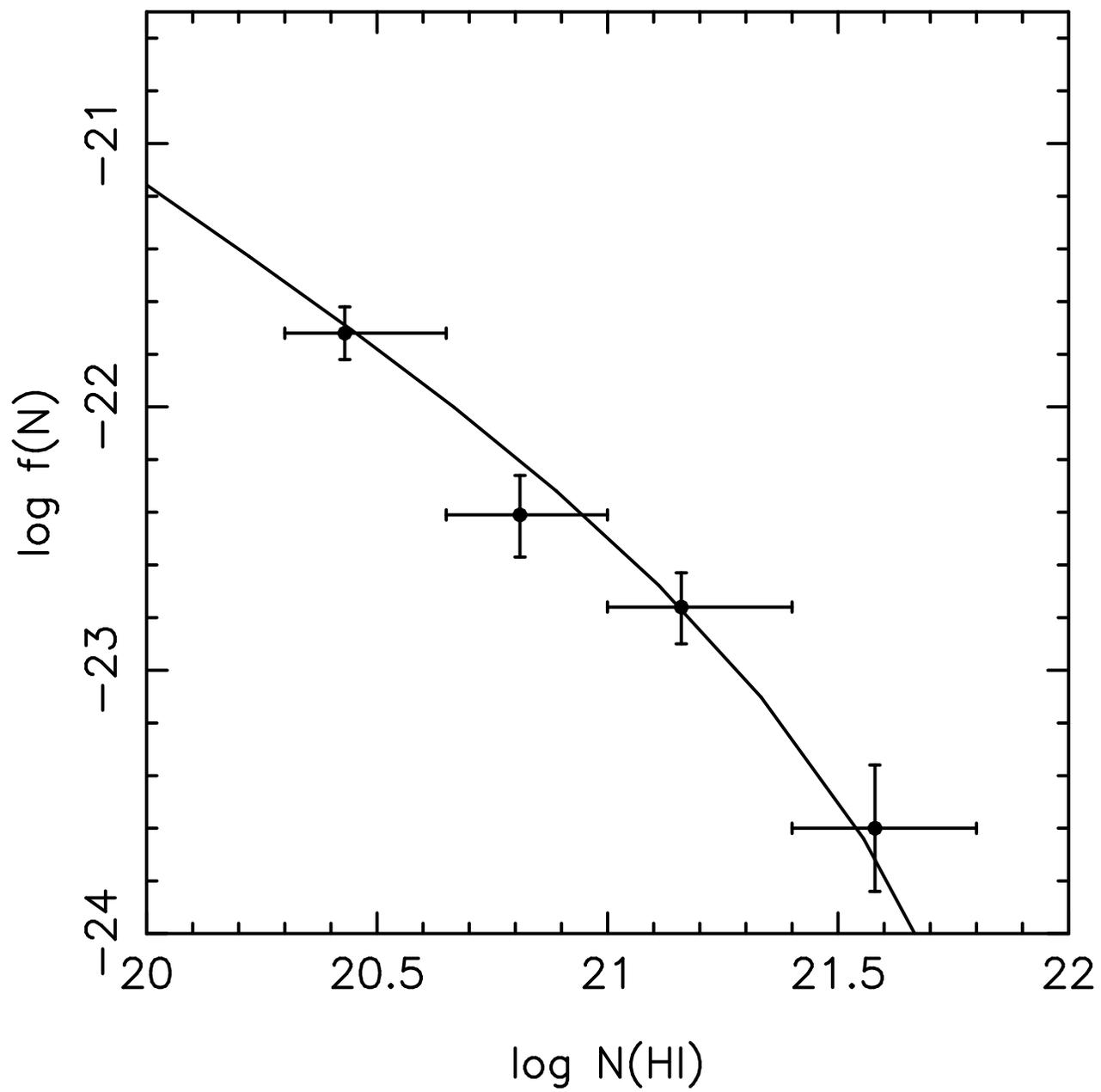
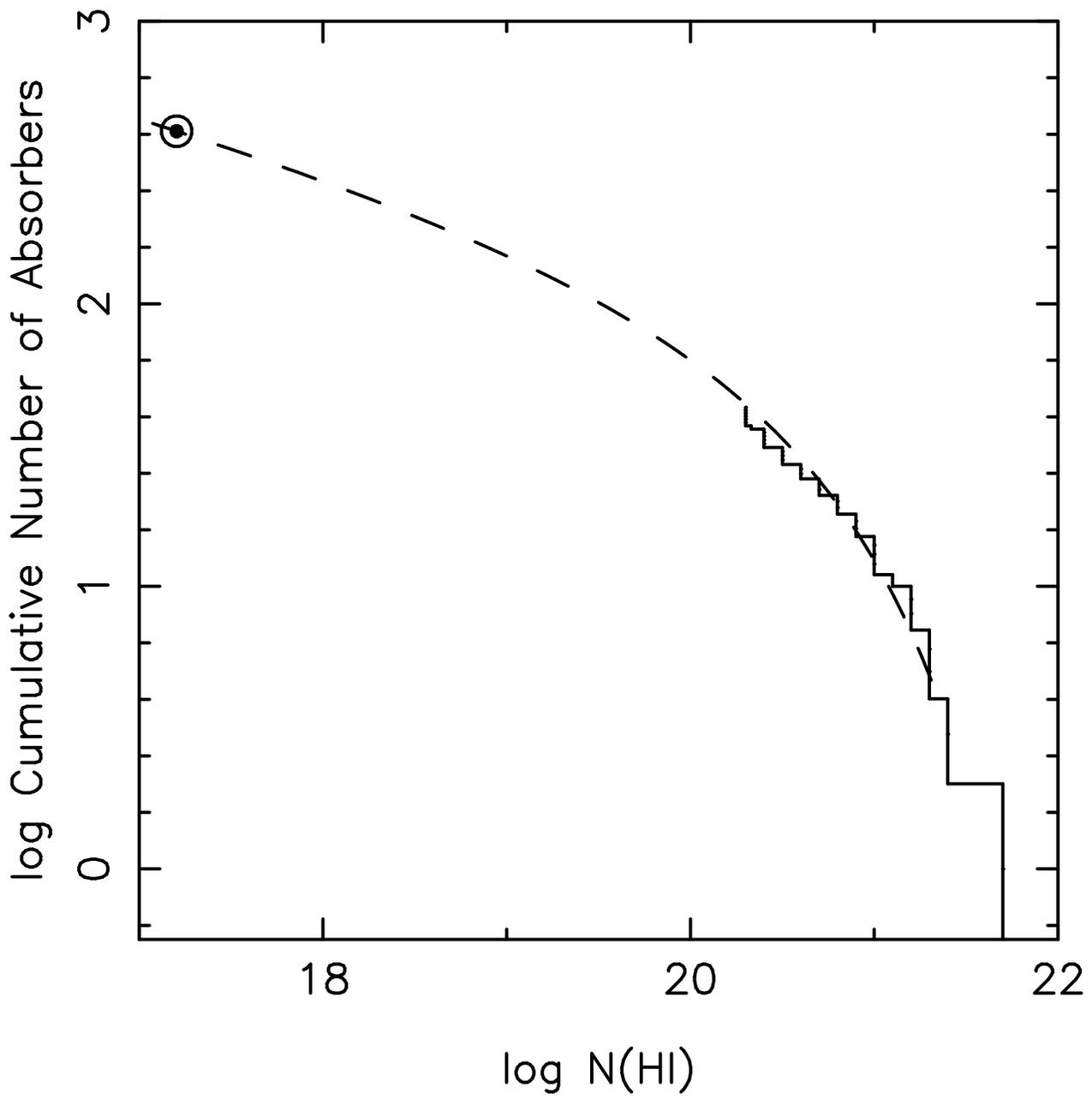
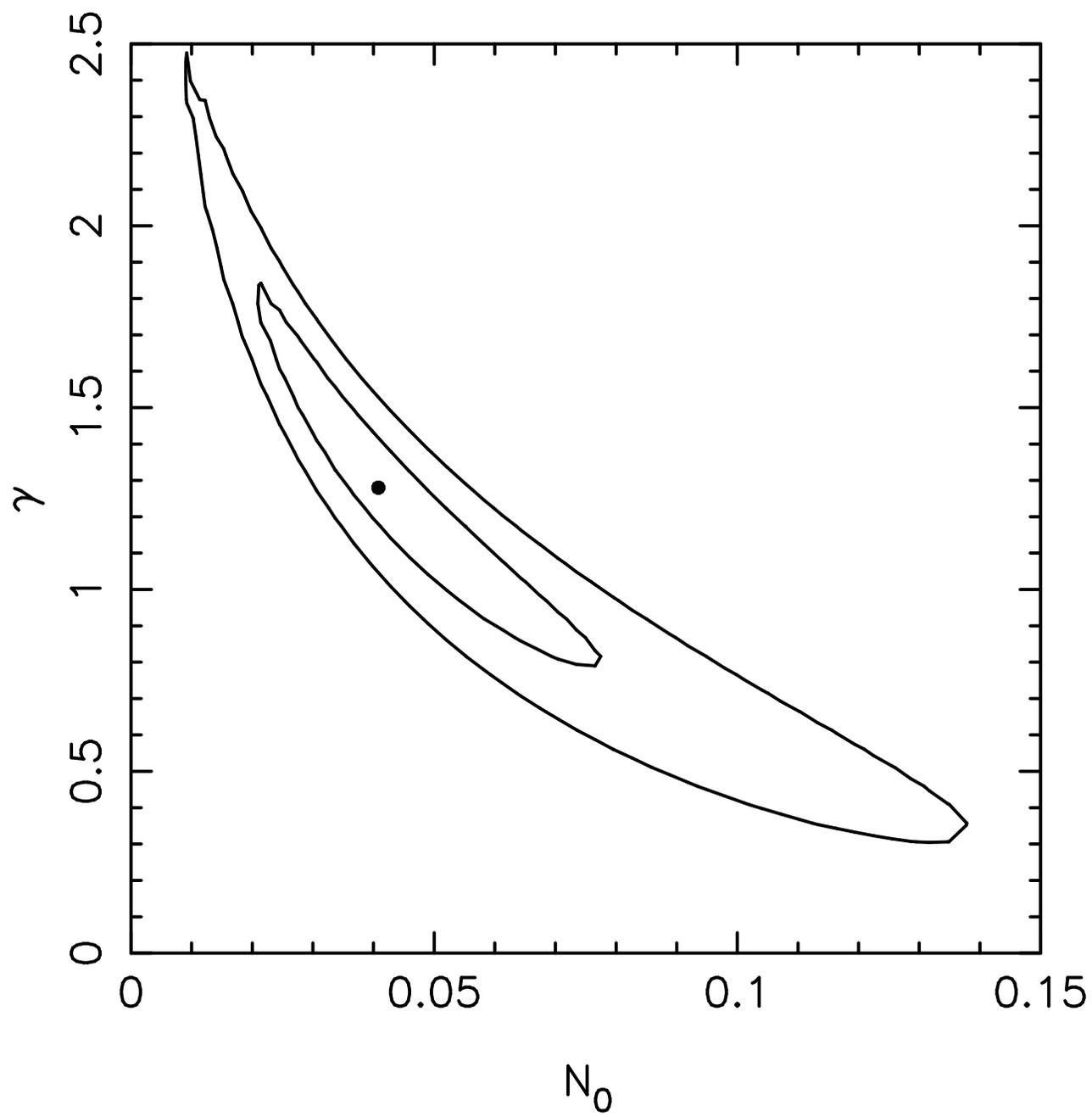
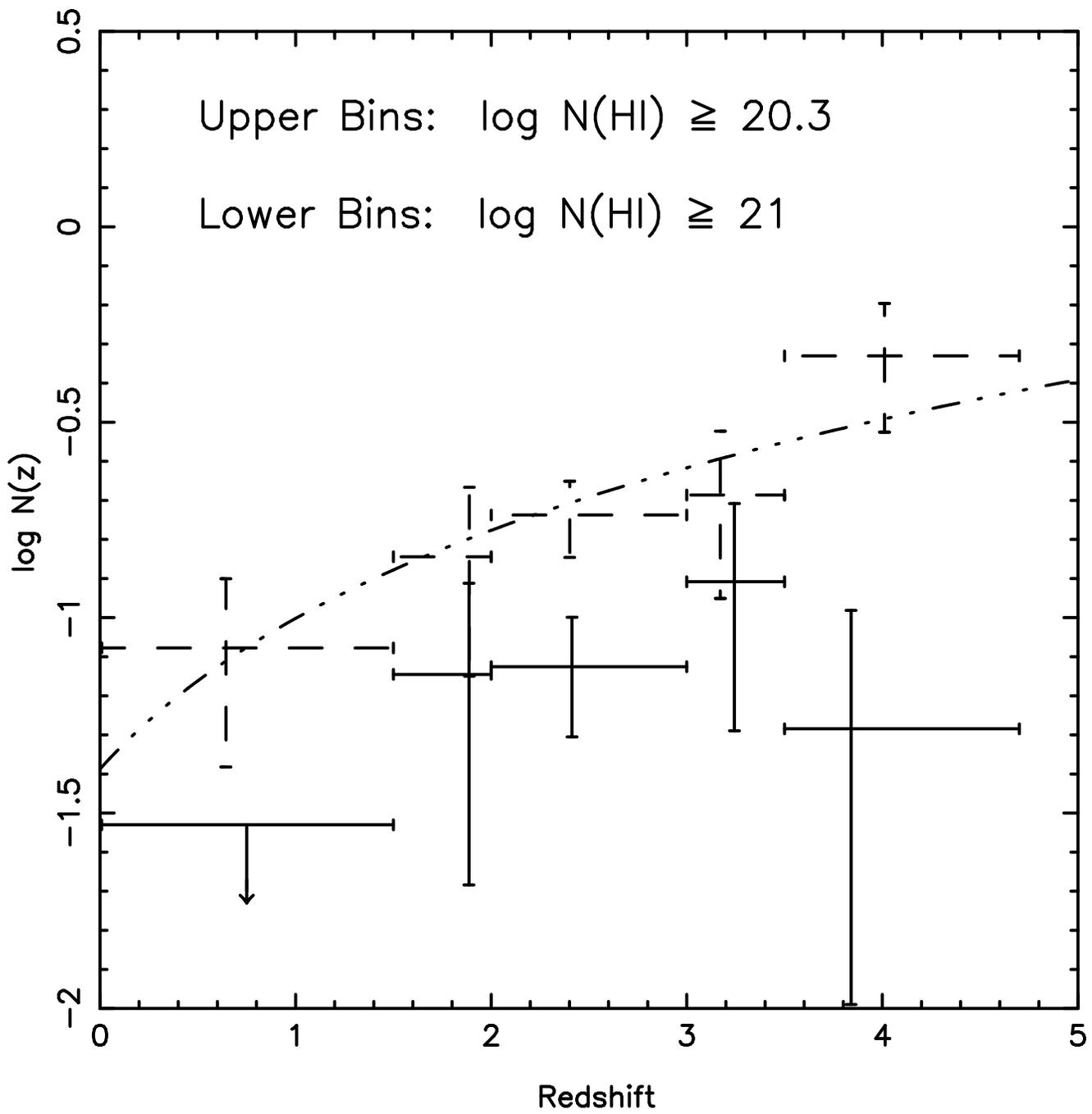
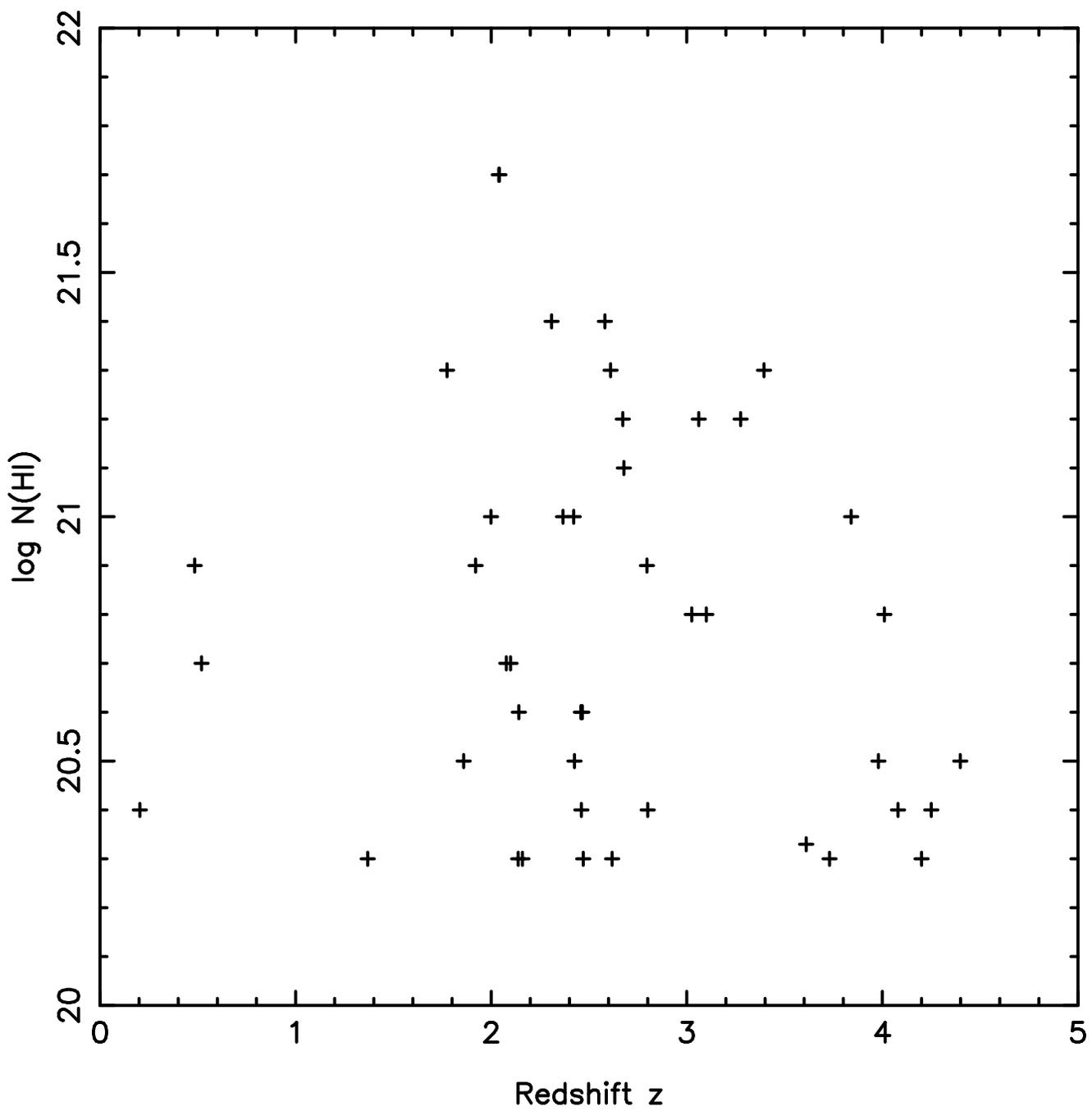
\end{document}